\documentclass{WileyMSP-template}
\usepackage{indentfirst} 
\usepackage{amsmath} 
\usepackage{ragged2e}
\usepackage{cite}
\begin{document}

\pagestyle{fancy}
\rhead{\includegraphics[width=2.5cm]{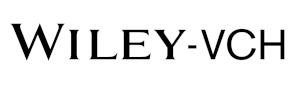}}

\title{A high-Q metasurface signal isolator for 1.5T surface coil magnetic resonance imaging on the go}

\maketitle

\author{Qun Ren$^{1,3}$}
\author{Yuxin Lang$^{1}$}
\author{Yuqi Ja$^{1}$}
\author{Xia Xiao$^{2}$}
\author{Yu Liu$^{2}$}
\author{Xiangzheng Kong$^{2}$}
\author{Ruiqi Jin$^{1}$}
\author{Yongqing He$^{1}$}
\author{Jianwei You$^{3}$}
\author{Wei Sha$^{4}$}
\author{Yanwei Pang$^{1}$}

\begin{affiliations}
$^{1}$School of Electrical and Information Engineering, Tianjin University, Tianjin 300072, China
\end{affiliations}
\begin{affiliations}
$^{2}$School of Microelectronics, Tianjin University, Tianjin 300072, China
\end{affiliations}
\begin{affiliations}
$^{3}$State Key Laboratory of Millimeter Waves, School of Information Science and Engineering, Southeast University, Nanjing, 210096, China
\end{affiliations}
\begin{affiliations}
$^{4}$Key Laboratory of Micro-Nano Electronic Devices and Smart Systems of Zhejiang Province, College of Information Science and Electronic Engineering, Zhejiang University, Hangzhou, 310027, China
\end{affiliations}
\begin{affiliations}
Email Address: pyw@tju.edu.cn
\end{affiliations}



\justifying
\begin{abstract}
The combination of surface coils and metamaterials remarkably enhance magnetic resonance imaging (MRI) performance for significant local staging flexibility. However, due to the coupling in between, impeded signal-to-noise ratio (SNR) and low-contrast resolution, further hamper the future growth in clinical MRI. In this paper, we propose a high-Q metasurface decoupling isolator fueled by topological LC loops for 1.5T surface coil MRI system, increasing the magnetic field up to fivefold at 63.8 MHz. We have employed a polarization conversion mechanism to effectively eliminate the coupling between the MRI metamaterial and the radio frequency (RF) surface transmitter-receiver coils. Furthermore, a high-Q metasurface isolator was achieved by taking advantage of bound states in the continuum (BIC) for extremely high-field MRI and spectroscopy. An equivalent physical model of the miniaturized metasurface design was put forward through LC circuit analysis. This study opens up a promising route for the easy-to-use and portable surface coil MRI scanners.
\end{abstract}


\section{Introduction}
 Magnetic resonance imaging (MRI) is widely used in the early diagnosis and screening of systemic tumors and nervous system diseases, due to its noninvasive and high-penetration characteristics.\textsuperscript{\cite{ref1}} This powerful imaging technology has growing in popularity in clinical devices on the go, becoming an essential tool in early diagnosis and screening of systemic tumors, nervous system, cardiovascular system and other diseases.\textsuperscript{\cite{ref2,ref3,4,5}} Specifically, signal-to-noise ratio (SNR) is one of the important indexes to evaluate image quality. A higher SNR means higher spatial resolution and faster images, which is the key technology to facilitate future clinical MRI examinations. Therefore, it is of great significance to improve the SNR of MRI, which is possible to be achieved by intelligent magnetic metamaterials with very low cost of upgrading high-end power scanners.\textsuperscript{\cite{6,7,8,9}} Numerous previous studies have demonstrated the potential of metamaterials in enhancing the SNR of RF surface coils within MRI systems. Pursuing local surface coil MRI is an appealing alternative to traditional  birdcage coil transmission, especially for patients with implants, with substantially enhanced magnetic field, expanded imaging range and greatly improved SNR.\textsuperscript{\cite{10,11,12}}

However, it is inevitable that a certain coupling interaction occurs in between, resulting in reduced signal transmission efficiency, amplified noise and unexpected safety concerns.\textsuperscript{\cite{13,14}} Particulally, an additional magnetic field inhomogeneity might bring about the introduction of artifacts and even image failure.\textsuperscript{\cite{15,16,17}} Typically, a common method for decoupling to impede mutual interference between signals is through classical shielding.\textsuperscript{\cite{18,19,20}} However, the combination of surface coils and magnetic resonance imaging (MRI) signal enhancers based on metamaterials is not suitable for this decoupling approach because what we require is achieving simultaneous signal superposition and enhancement between the two while maintaining isolation.\textsuperscript{\cite{21,22,23,24}} Indeed, balancing decoupling and magnetic resonance imaging (MRI) signal enhancement has become an important technical challenge. Achieving both objectives simultaneously requires careful consideration and design to optimize the performance of the system.\textsuperscript{\cite{25,26,27}}

Here, we report a high-Q metasurface-based decoupling isolator for MRI fueled by surface coils to effectively eliminate this coupling through polarization conversion. While different schemes to manipulate and enhance magnetic fields have been presented, we propose a configuration of a multi-layer decoupling isolator and demonstrate the practicality and efficacy of utilizing it for friendly clinical applications. Moreover, we examine the principles of cross-polarization conversion to clarify the mechanism of the isolator when expose to incident circular-polarized plane waves. Furthermore, by breaking the symmetry of the metal winding structure, we present the potential to achieve high Q with bound states in the continuum. Lastly, the performance of the decoupling isolator is testified with wrist and leg imaging as examples. As a result, the resonant mode of the metamaterial is excited during the MRI procedure, leading to a marked boost in magnetic field strength.

\section{Design of decoupling metasurface isolator}

The decoupled metasurface isolator excited by circular-polarized signal designed in this paper consists of three cascading layers of unit cell arrays as shown in Fig. 1(a), with length of 29.48 mm, width of 23.4 mm, and height of 58 mm, respectively. The metal winding ($w_{r}$= 40.5 mm, $g_{d}$= 36 mm and $w_{d}$= 3 mm), was fabricated on both sides of each dielectric substrate, connected with the cuboid metal strips, to realize cross polarization transformation which is essential for decoupling. 

\begin{figure}[h]
  \includegraphics[width=\linewidth]{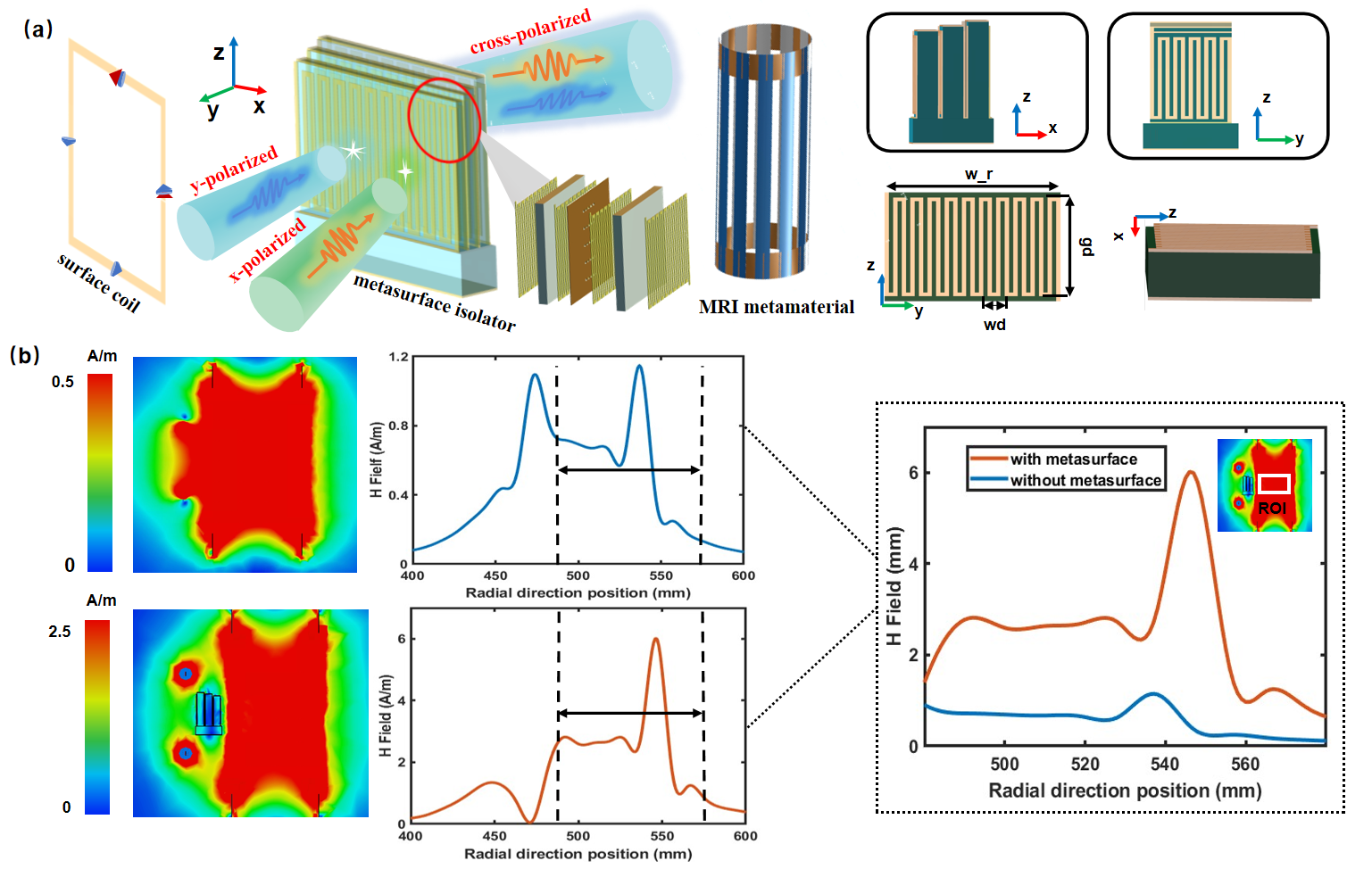}
  \caption{(a) Co-simulation diagram of the MRI enhancement system, with 3-view drawings of decoupling metasurface isolator constructed by topological loops on the right. The metasurface can be divided into three layers, including three dielectric substrates with varying widths (42 mm, 0.95*42 mm, 0.9*42 mm), along with metal winding wires on both sides of each substrate, placed at a uniform distance of 1.7 mm. (b) Magnetic field enhancement and the radial magnetic field strength distribution spectra with/without the metasurface isolator.}
  \label{fig:boat1}
\end{figure}

In order to explore the decoupling effect and magnetic resonance enhancement of the decoupling isolator, in the following discussion, a topological circuit model of the decoupling process was analyzed via joint simulation of the surface coil and metamaterial. Herein to simplify the process, the surface coil was constructed using a single square metal ring interspersed with capacitors, each featuring 4 evenly spaced notches by 90° interval. An adjustable capacitor was strategically placed within each gap for tuning frequencies. In the meantime, the metamaterial consisted of 12 strips, each separated by 30° angle. Each strip adopted a three-layer structure, comprising a metal coating at the bottom, a dielectric material in the middle, and metal rings at both ends serving as fixtures and connectors. 

Our 1.5T magnetic resonance imaging system was excited by a circular polarized signal, which can be decomposed into two distinct ports with 90° phase difference. A magnetic field probe was placed in the center of the metamaterial at radial coordinate of 530 mm. The magnetic field enhancement and the radial distribution of magnetic field, was presented in Fig.  1(b), within the region of interest (ROI), particularly with radial distribution along Z coordinate. The side view of the magnetic field between the surface coil and metamaterial indicates coupling between these two components, resulting in a reduction of overall magnetic field. Keeping the spacing between the metamaterial and surface coil unchanged, when decoupling metasurface isolator was put in between, the magnetic field enhancement operated at 63.8 MHz can be obviously improved.

It's worth noting that in the absence of decoupling isolator, the magnetic field spectrum exhibits two peaks in the radial direction (as shown in Fig. 1(b) above, at 470mm and 540mm). However, upon introducing the decoupling isolator, the first peak vanishes (as depicted in figure 1(b) below), leaving only peak at 540mm. This happens because, in spite of the decoupling isolator there is still some degree of coupling with the metamaterial itself. As a result, there is a reduction in magnetic field strength near the isolator (at 470 mm). Within the inner ROI in Fig. 1(b), as the radial distance increases, both curves consistently demonstrate the same trend, further confirming the decoupling effect achieved by isolator.

The presence of the decoupled metasurface within the metamaterial results in a magnetic field enhancement by 4-6 times, where the most significant enhancement occurs at a radial distance of 550 mm. Consequently, in the joint simulation of all 3 components, the decoupling process between RF coil and MRI metamaterial brings about a collaborative enhancement of the magnetic resonance imaging process.

\section{Discussion}
\subsection{Equivalent topological LC circuit of the metasurface isolator}

To analyze the electromagnetic behavior of the metal-dielectric hybrid metamaterial, the overall structure can be decomposed into 3 cascaded layers, represented as 'Part1', 'Part2', and 'Part3' in an equivalent circuit, as shown in figure 2. An equivalent LC circuit model is established based on the structure of the metasurface isolator. The base layer consists of a bottom metal square patch coupled to a dielectric layer of substrate. This layer not only reflects incident electromagnetic waves but also exhibits bandstop characteristics. On basis of above nature, further winding in the dielectric layer and the adjacent serpentine metal winding structure exhibit bandpass characteristics determined by their slotted design, which can be represented as a parallel combination of inductors and capacitors. The serpentine metal structure sandwiched on both sides can be seen as forming a connected configuration in series. Hereby, for the single-side metal structure, the connection between the two side metal structures is established through upper and lower metal strips to form a series circuit referred to as 'Part1', and so on.

\begin{figure}
\centering\includegraphics[width=0.8\textwidth]{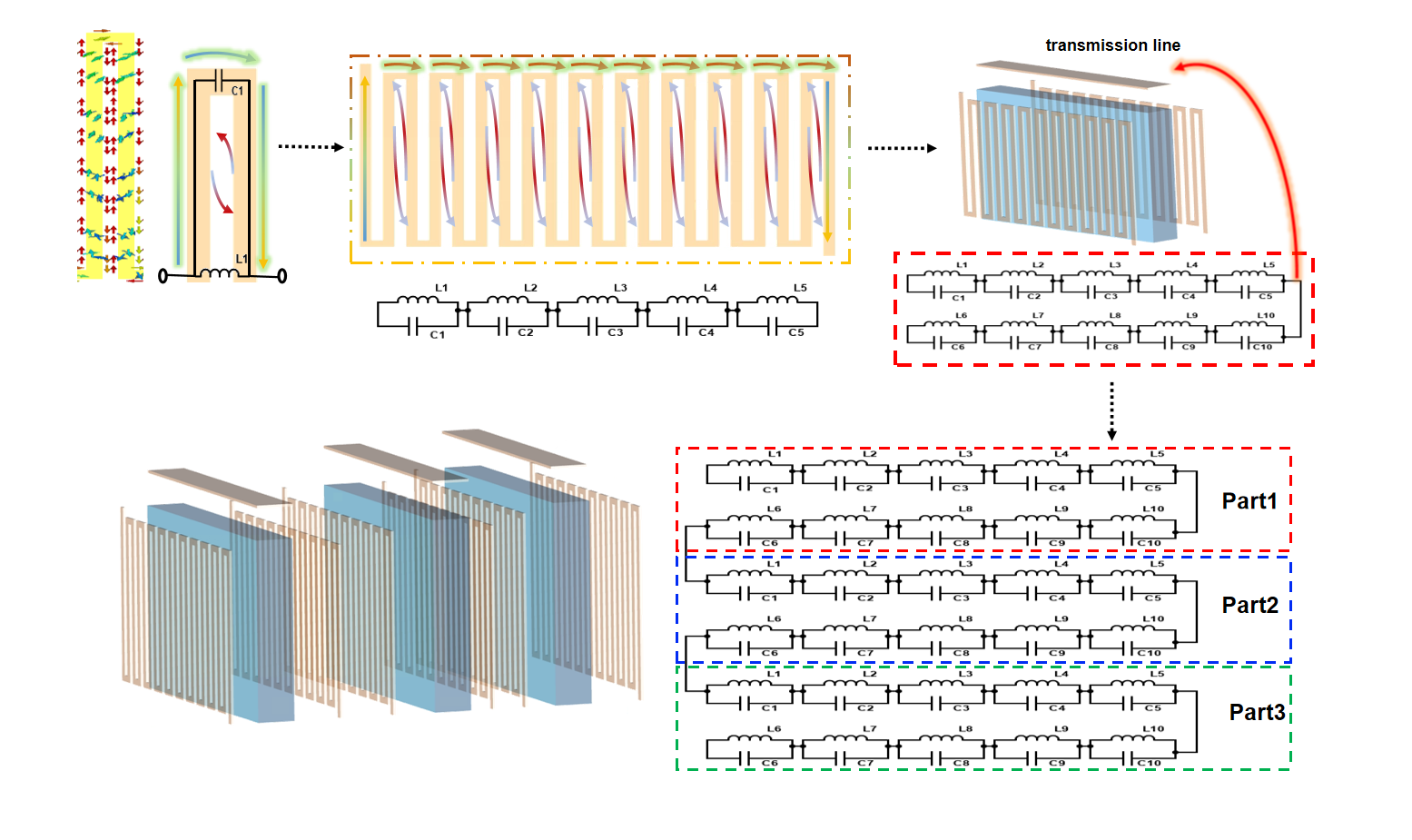} \caption{ The derivation of an equivalent LC circuit schematic for a three-layer cascaded decoupled metasurfaces. The upper left is the current distribution and circuit model of a single-winding structure, the middle is the current distribution and circuit model of the single-layer metal winding structure, and the upper right is the current distribution and circuit model of the double-layer metal winding structure after with the addition of the dielectric medium. Below is the breakdown diagram of the three-level cascaded structure and corresponding circuit model.}\label{fig:boat1}
\end{figure}

The basic principle of the isolator resonance performance is related to the L-C circuit resonance and resonant frequency, and the resonant frequency $ f=\frac{1}{2\pi\sqrt{LC}} $ , where L and C are the equivalent inductance and capacitance, respectively. Inductance L includes the self inductance and mutual inductance of array elements and rings.\textsuperscript{\cite{28}} In the multilayer structure, we analyze each layer of isolator individually to obtain the equivalent circuit of each layer structure and cascade it through the transmission lines.\textsuperscript{\cite{29}} According to the current distribution and electric field distribution of each layer of isolator, an equivalent circuit model of metal-dielectric stacked metamaterials is proposed. Each loop of the dielectric layer and the serpentine metal winding structure on both sides provides bandpass characteristics due to its slotted structure, so it is modeled by a parallel combination of inductance and capacitance.\textsuperscript{\cite{30}} The serpentine metal structure on each side can be seen as a series form of each loop, so the single-sided metal structure is modeled by series. The metal structure on both sides is connected by the top metal on top and the lower part is connected by metal strips, so it is essentially a series circuit, which is regarded as Part 1. In this way, a total of three layers are cascaded, and the corresponding circuit model is decomposed into series circuits of Part1, Part2, and Part3. In absence of the other metal layers, the square metal patch at the bottom layer consists of metal square patches that reflect incident electromagnetic waves and thus block the electromagnetic wave.\textsuperscript{\cite{31}} Therefore, it is modeled by a series combination of inductors and capacitors to provide band-resistance characteristics. The value of $C_{bottom}$ for square patches of length L and metal patches spacing h is given by:
\begin{equation}
C_{bottom}=\varepsilon _{0}\varepsilon _{eff}\frac{2Llog(cosec\frac{\pi h}{2L})}{\pi},
\end{equation}
where $\varepsilon _{0}$ is the vacuum permittivity, $\varepsilon _{eff} = 0.5(1+\varepsilon_{r})$ , $\varepsilon _{r}$ is the permittivity of the dielectric layer, L is the length of the square patch, h is the distance from the bottom plate to the dielectric layer. The inductance $L _{bottom}$ of the square metal patch backed by a single layered dielectric substrate of permittivity $\varepsilon _{r}$ is given by:
\begin{equation}
\omega _{0}=\frac{1}{2\pi \sqrt{L_{bottom}C_{bottom}}}.
\end{equation}

For a three-level structure, the capacitance solution formula for each parallel circuit is:
\begin{equation}
C=\frac{\varepsilon _{eff}(N-1)D}{18\pi}\frac{K(k)}{K'(k)},
\end{equation}
where $K(k)$ is the complete elliptic integral of the first kind and its complement is $K'(k)$, D is the length of the metal bar of the serpentine metal structure, N is the number of metal bars.

Herein,
\begin{equation}
\frac{K(k)}{K'(k)}=\left\{\begin{array}{rcl}
    \frac{1}{\pi}ln[\frac{2(1+\sqrt{k})}{1-\sqrt{k}}] & \mbox{for} & 0.707\leq k\leq 1 \\
    \frac{\pi}{ln[2(1+\sqrt{k'})(1-\sqrt{k'})]} & \mbox{for} & 0\leq k\leq 0.707
     \end{array}\right
.\end{equation}

and
\begin{equation}
k=tan^2(\frac{a\pi}{4b}), a=\frac{w_{d}}{2}, b=\frac{w_{d}+g_{d}}{2}, k'=\sqrt{1-k^2}
\end{equation}
where $w_{d}$ is the width of each metal winding and $g_{d}$ is the length of the metal-winding wire.

The inductance continues to be solved using equation (2). Finally, the equivalent capacitance of the entire circuit model is 89pF and the equivalent inductance is 70nH, which meets the requirement of 63.8MHz resonance frequency in 1.5T magnetic resonance system.

\subsection{Underlying physics of decoupling process}

After the incident x-polarized and y-polarized waves pass through the decoupling isolator, the transmitted waves become orthogonal waves (y-polarized and x-polarized waves, respectively), so as to reduce the coupling between surface coil and MRI metamaterial and superimpose the mode vectors of both. According to the polarization conversion principle of metasurface isolator, the electric field distribution of the metal winding structure during the conversion of x-polarized wave and y-polarized wave is further analyzed.\textsuperscript{\cite{32}} The principle analysis is shown in Fig. 3(a). The uv coordinate system is the relative coordinate system obtained through rotating the xy coordinate system by 45 degree counterclockwise to resolve the incident electromagnetic wave into the uv direction.\textsuperscript{\cite{33}} The polarization rotation of transmitted electromagnetic wave is easy to observe then in this way. Specifically, the incident electromagnetic wave can be expressed by the following formula:
\begin{equation}
\boldsymbol{E} _{i}=\boldsymbol{v}\boldsymbol{E}_{iv}e^{j\varphi_{iv}}\boldsymbol+{u}\boldsymbol{E}_{iu}e^{j\varphi_{iu}}
\end{equation}
where $\boldsymbol{E} _{i}$ represents the incident linear polarized wave, $\boldsymbol{E}_{iv}$ and $\boldsymbol{E}_{iu}$ indicate the incident wave amplitude in the v and u direction, and $\boldsymbol{v}$,$\boldsymbol{u}$ indicate the unit vector in the v and u directions, respectively. $e^{j\varphi_{iu}}$ and $e^{j\varphi_{iv}}$ are the phase of the two incident wave components, respectively.

The transmitted wave can be expressed as follows:
\begin{equation}
\boldsymbol{E} _{t}=\boldsymbol{v}\boldsymbol{t}_{v}\boldsymbol{E}_{tv}e^{j\varphi_{tv}}\boldsymbol+{u}\boldsymbol{t}_{u}\boldsymbol{E}_{tu}e^{j\varphi_{tu}}
\end{equation}
In Eq. (7), $\boldsymbol{E} _{t}$ is the transmitted wave,  $\boldsymbol{t}_{v}$ and $\boldsymbol{t}_{u}$ represents the transmission efficiency of the electric field components in both directions, $e^{j\varphi_{tv}}$ and $e^{j\varphi_{tu}}$ are the phases of these two transmitted wave components, respectively.

\begin{figure}[h]
\centering\includegraphics[width=0.9\textwidth]{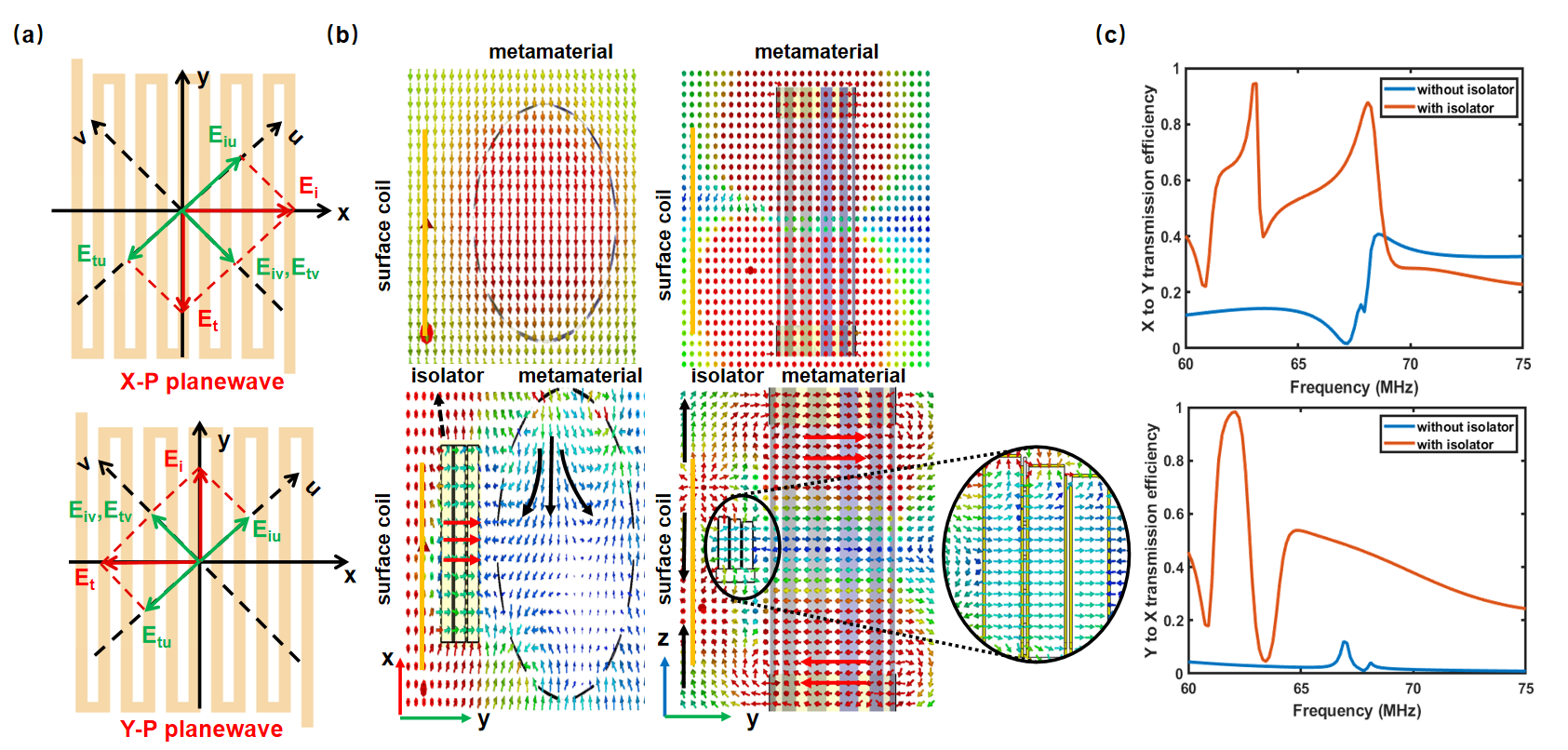} \caption{(a) Decomposition of electromagnetic waves on the a single layer serpentine metal structure in the uv coordinate system, where the left incident wave is x-polarized and the right one is y-polarized. (b) Schematic diagram of the current direction. The top is the result when no isolator is added. It can be seen that the current direction on the surface coil and MRI metamaterial is the same in the x direction. Below is the result of adding the isolator. The current on the surface coil is in z direction, whereas the current on the metamaterial is in y direction, so that the electric field becomes orthogonal by passing the isolator. This proves the orthogonal polarization conversion process of the isolator. (c) Polarization conversion ratio (PCR) result in the MRI operating band. Above is the conversion efficiency of x-polarized wave to y-polarized wave, whereas below is the conversion efficiency of y-polarized wave to x-polarized wave.}\label{fig:boat1}
\end{figure}

As shown in Fig. 3(b), the decomposition of the electric field, when the incident wave incident along the x-axis direction, because of the presence of the serpentine metal winding vertical bars on the left and right sides, the electric field component in the u direction changes in direction 180°, if $\boldsymbol{t}_{u}\approx \boldsymbol{t}_{v}$ at this time, the direction of the resultant transmission wave direction is the y-axis direction, which makes the $\boldsymbol{E} _{t}$ linear polarization wave conversion to design cross polarization.\textsuperscript{\cite{34}} When the incident wave is y-polarized wave, can be found that the unit structure along the y axis is positive direction and negative direction are two metal structure, so the electromagnetic wave and metal structure coupling, decomposition of the component on the uv direction, u direction of the electric field also of 180 direction change, combined behind the analysis, can find y-polarized wave incident, electromagnetic wave basically all the cross transmission transition.\textsuperscript{\cite{35,36}} The case shown in Fig. 3(c) is the simulation result obtained from the negative direction of the incident wave, that is, the vertical incident from the front of the metasurface, which is divided into two cases: the incident wave is x-polarized (PCR1) and y-polarized (PCR2). The PCR in figure can be defined as:

\begin{equation}
PCR = \vert Txy\vert^2/(\vert Txy\vert^2+\vert Tyy\vert^2)
\end{equation}
$Txx$ and $Tyy$ are transmission coefficients of isopolarization under the front or reverse incident, respectively, $Txy$ and $Tyx$ are transmission coefficients of cross polarization under the front or reverse incident, respectively. To further understand the conversion physics of this multilayer structure, which means a 90° change will be achieved when the polarization direction of the incident EM wave is along the y-axis(or x-axis), we first define the polarization direction is along y-axis for instance, then the cross- and co-polarization can be depressed as $T_{xy}=\vert E_{xt}/E_{yi} \vert$ and $T_{yy} =\vert E_{yt}/E_{yi}\vert$ respectively, where E indicates the electric field; the subscript i and r defines the incidence and transmission of EM waves, respectively; and the subscripts x and y denote the polarization directions of EM waves, respectively.\textsuperscript{\cite{13,14}}

To assess the polarization transformation effect of the proposed isolator, we conducted a comprehensive analysis through both electric field characterization and the calculation of the PCR. In Fig. 3(b), we present the electric field distribution map, providing top and side views. In the top view, shown on the left, we present the results without incorporating the isolator, demonstrating that the electric field consistently aligns with the x-axis. This consistency holds true on both the surface coil side and within the metamaterials region. Conversely, in the bottom view, after introducing the isolator, a conspicuous orthogonal polarization conversion effect becomes evident. The electric field on the surface coil shifts towards the z-axis, while significant orthogonal changes occur post-isolator integration, resulting in a shift towards the y-axis. Additionally, the electric field within the metamaterial now aligns with the isolator direction, extending along the y-axis. This observation leads us to conclude that the isolator effectively decouples the system by orthogonal electric field direction conversion. This conversion establishes orthogonality between the electric field orientations on the surface coil side and the metamaterial side, resulting in a substantial reduction in coupling losses and a notable enhancement in polarization conversion efficiency.

Figure 3(c) illustrates the PCR as a function of frequency. The upper section represents the conversion efficiency of x-polarized waves to y-polarized waves, while the lower section illustrates the conversion efficiency of y-polarized waves to x-polarized waves. Remarkably, at a frequency of 63.8 MHz, the PCR calculated after incorporating the isolator consistently exceeds 80$\%$, in stark contrast to the PCR observed without the isolator, which remains low, typically below 20$\%$, which underscores the remarkable performance enhancement achieved with the isolator integration. In essence, this indicates that over the extensive frequency range, both x-polarized and y-polarized incident electromagnetic waves consistently exhibit an approximate 90-degree phase shift, affirming the isolator effectiveness in achieving polarization orthogonality.

\subsection{Realization of ultra high Q factors}

In this section, symmetry-protected BIC is formed through broken the structural symmetry within the system to achieve high Q factor. When a state exhibits different symmetry characteristics compared to its surroundings, it is decoupled from radiation and behaves as BIC. This unique state, known as a symmetry-protected BIC, maintains its localization as long as the system symmetry remains unchanged.\textsuperscript{\cite{37,38}} The distinguishing feature of symmetry-protected BICs is their predictability, which avoid fine-tune structural parameters in wave vector space to locate them, which have been extensively studied across various systems, including metasurfaces, sub-wavelength structures, classical gratings, and photonic crystals.\textsuperscript{\cite{39,40}}

Since BICs represent discrete modes effectively decoupled from the radiation continuum of different symmetric classes, when these discrete modes are excited in an asymmetric manner, their electromagnetic field distributions deviate from the symmetry mode, leading to the emergence of ultra-high Q resonances. The extent of leakage resonance depends on the degree of structural asymmetry.\textsuperscript{\cite{41}} By introducing periodic perturbations in the direction or position of adjacent elements, in-plane symmetry can be broken, even at oblique or positive incidence angles, allowing these modes to leak into free space.\textsuperscript{\cite{42}}

\begin{figure}
\centering\includegraphics[width=1\textwidth]{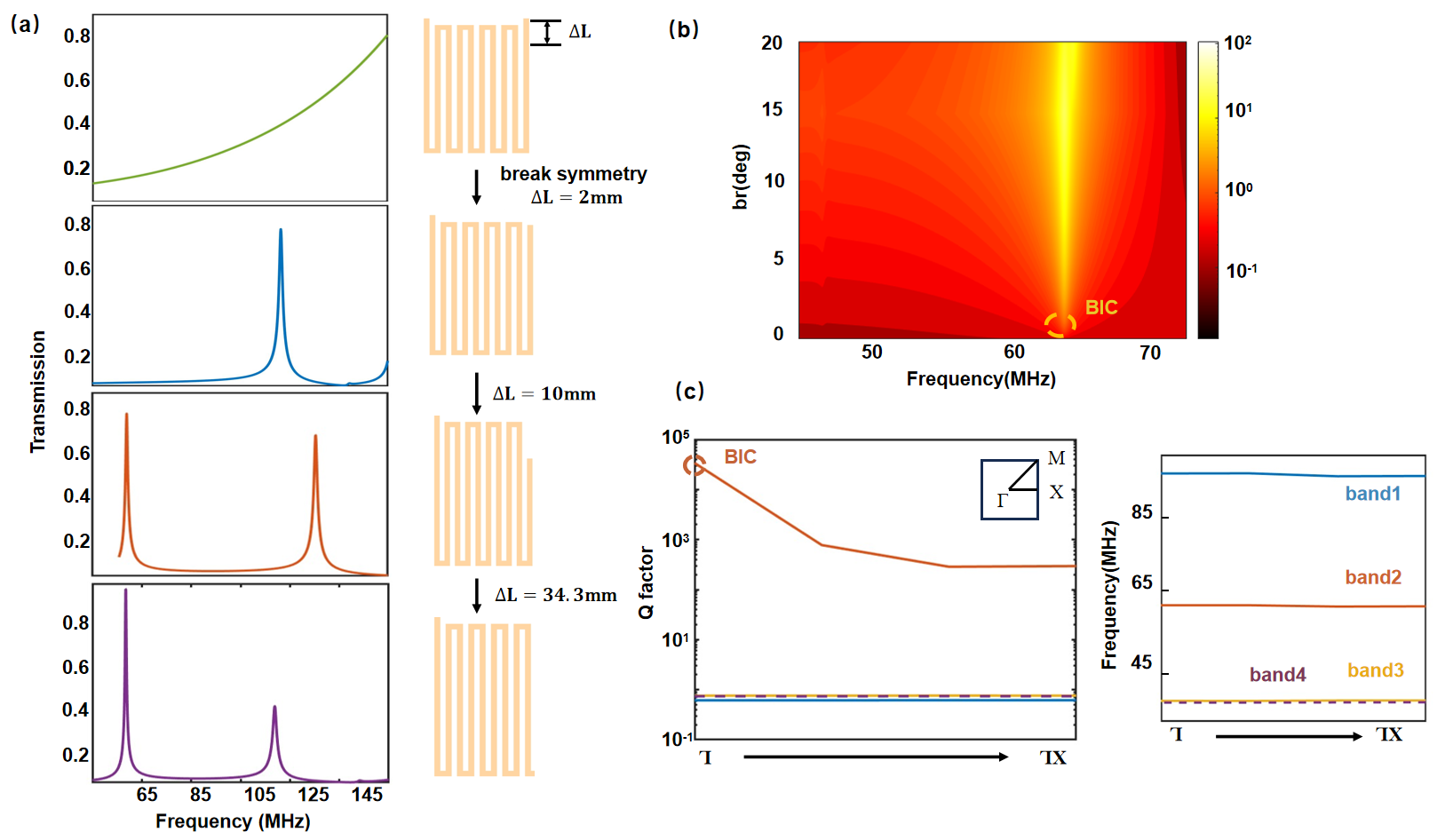} \caption{(a) Transmission spectra for \ \(\Delta L\)= 0 mm, 2 mm and 10 mm. (b) Discrete spectra and color plot of simulated transmission spectra with different \ \(\Delta L\), changing \ \(\Delta L\) from 0 mm to 20 mm. (c) Corresponding Q factor and band structure for the BIC.}\label{fig:boat1}
\end{figure}

Hereby, we introduce a metasurface achieving a high Q-factor BIC in the cross-polarization configuration. Initially, in the case of symmetry on the left and right sides of the metal winding structure, there is no resonant peak in the spectral response. Then, we break the symmetry of the structure and reduce the length of the leftmost metal strip, and the amplitude of the reduction, that is, the length difference between the left and right metal bars, is expressed by \(\Delta L\). When the length of the right metal strip is reduced to 1 mm, the transmittance decreases significantly at 140 MHz and an unexpected quasi-BIC is excited, regarded as BIC1. However, the BIC1 frequency does not meet the frequency band requirement of 1.5T magnetic resonance system. Continue to reduce the length of the right metal bar, and when the gap increases to 8mm, a clear transmission valley can be observed on the transmission spectrum of 63.8 MHz, i.e., BIC2. In this case, BIC1 moves to lower frequency. Symmetry-protected BIC can be verified by breaking the structural symmetry of the device. We demonstrate the quasi-BIC phenomenon caused by symmetry fracture of metal winding structure, which has 6 different degrees of fracture, and the asymmetric transition from symmetry-protected BIC to quasi-BIC can be observed from the simulated transmission spectrum in the figure. The spectral properties of quasi-BICs also depend on the size of the symmetry break \(\Delta L\).

The disappearance of the resonance of the symmetric recovery region proves symmetry-protected BIC, where the figure of merit tends to infinity as the symmetrical gap of the broken symmetry becomes zero. The transmission response of metal wound structures with different clearances (\ \(\Delta L\)) was scanned. When the symmetry is broken, a clear quasi-BIC phenomenon can be seen at \ \(\Delta L\)=8mm. For an ideal BIC, the theoretical merit factor should be close to infinity when the imaginary part of the permittivity is zero. However, the inherent losses in metal lead to limited quality factors in depleted BICs. Figure 4(c) shows the Q factor calculated by this method to break the symmetry and symmetric structure at about 63.8 MHz, and shows that the Q factor of the quasi-BIC calculated at the symmetry point rises rapidly, indicating that the point is aligned with the protection of the BIC. The right of Fig. 4(c) shows a possible characteristic pattern at 63.8MHz in the \ $\Gamma -\Gamma X$ direction.


\section{MRI at 1.5T for a set of body phantoms}

A set of body phantoms representing various human tissue properties are simulated at 1.5T MRI condition. The dimensions of the phantom were 27 mm × 25 mm × 480 mm. The relative permittivity ($\varepsilon _{r} $) of the phantom is 68.4, and its conductivity ($\sigma $) is 0.4 S/m.\textsuperscript{\cite{43}} An RF surface coil is employed to generate a circularly polarized magnetic field (B1). Two excitation sources with a 90° phase difference are placed at positions differing by 90°, applying their excitation at the Larmor frequency in a counter-parallel and perpendicular manner to the static magnetic field.

\begin{figure}[htbp]
\centering\includegraphics[width=1\textwidth]{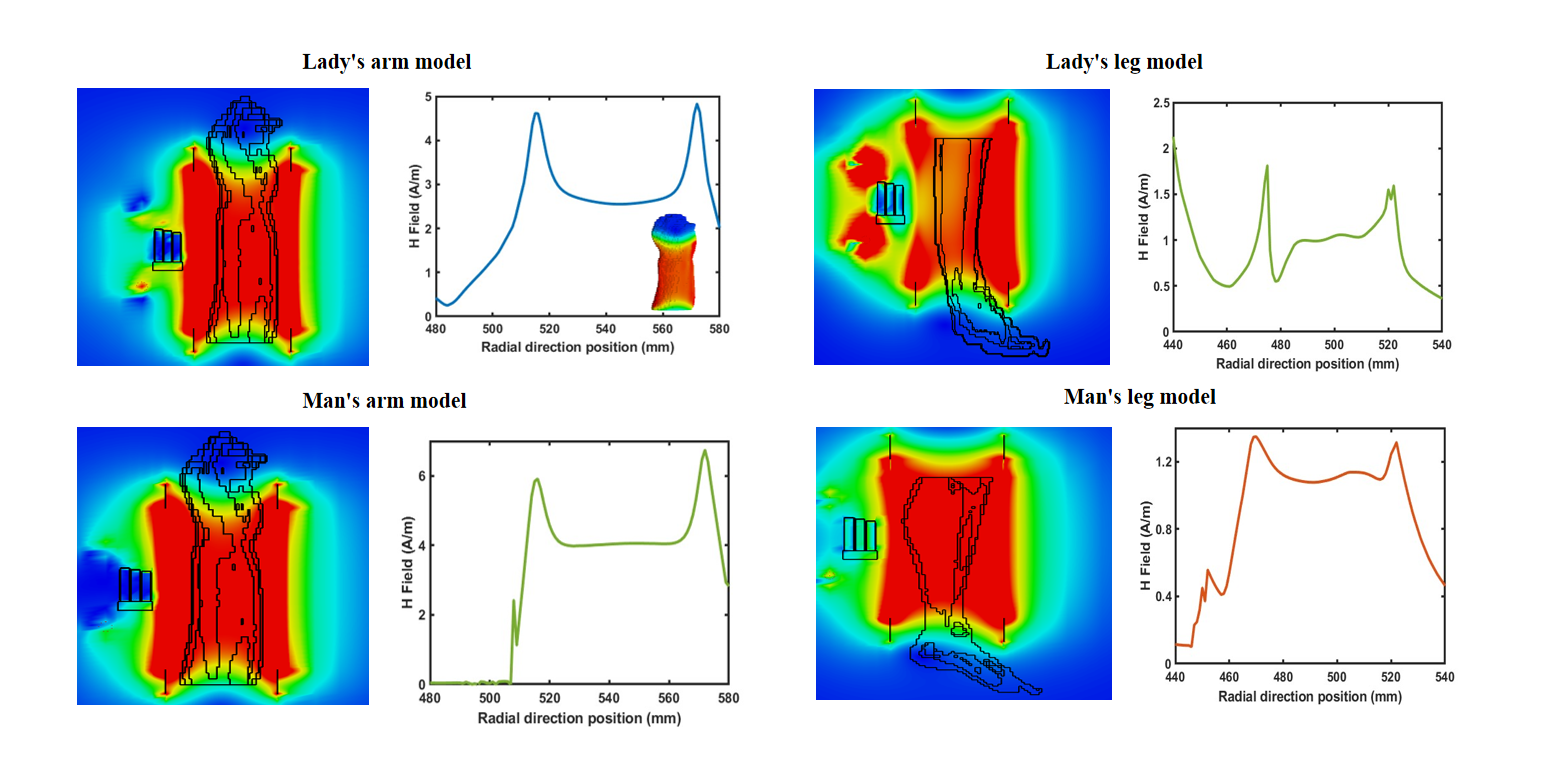} \caption{Magnetic field enhancement diagram and the magnetic field distribution with radial distance after adding the mannequin (arm). The baseline of the 2D magnetic field enhancement rendering on the left is set at 2.5 A/m.}\label{fig:boat1}
\end{figure}

The primary function of the surface coil in the MRI system is to excite unpaired proton spins of hydrogen atoms at the Larmor frequency during the transmit mode and to receive signals emitted by these protons at the same Larmor frequency during the receive mode.\textsuperscript{\cite{44}} Following the reciprocity principle, a simplified square-shaped surface coil is designed to operate exclusively in the receive mode, as shown in Fig. 1(a). The coil consists of a square metal loop with 4 equally spaced 90° wide gaps, each housing a tunable capacitor of 24 pF to tune the surface coil for matching the spin's Larmor frequency. Copper traces with a width of 0.5 mm are used to excite the phantom at 63.8 MHz. The coil is placed at a distance of 50 mm from the metamaterial.

Due to resonance scattering of the RF field and spatial redistribution of the electromagnetic field near the metamaterial at its resonance frequency, magnetic field enhancement occurrs. This resonance frequency matches the Larmor frequency of 1.5T MRI. The incident electromagnetic field experiences macroscopic effects due to the interaction between the surface coil and the metamaterial. When the RF wave impinges on the metamaterial, induced currents in the metal patches of each unit cell are observed. At resonance, the decoupling metamaterial produces enhanced magnetic flux density in the region of interest within the metamaterial.

Due to the slight resonance frequency shift observed after introducing the phantom in joint simulations, we add two monitors at 63.8 MHz and the resonance frequency after including the phantom. These monitors are used to observe the magnetic field distribution curves and magnetic field enhancement effects at these two frequencies. Figure 5(a) depicts the frequency-dependent magnetic field enhancement curves before and after the inclusion of the phantom. It is evident that prior to adding the phantom, the field strength at the resonance frequency of 63.8 MHz is 10 A/m. However, after introducing the phantom, the resonance frequency shifts to 63.8 MHz, with a field strength of 8 A/m.  In the figure located to the left above Fig. 5(a), precisely at 63.8 MHz, the magnetic field enhancement, occurring during the receive phase after adding the phantom, is illustrated. The maximum magnetic field value within the phantom with the metamaterial remains consistent at 2.5 A/m, matching the scenario without the phantom as shown in Fig. 1. The decoupling metamaterial enhances the magnetic field during the receive phase, mitigating field inhomogeneity.

In the upper right of Fig. 5(b), the results of the radial distribution of magnetic field along the z-axis are presented. It is evident that the results align with the magnetic field enhancement effect observed when the decoupling metamaterial is included, as shown in the lower right of Fig. 1(b). Notably, the magnetic field at the surface of the phantom during transmission is enhanced by 4 times. In the lower left of Fig. 5(b), precisely at the resonance frequency of 63.8 MHz, the maximum magnetic field is 4 A/m, which is stronger compared to the field strength at 63.8 MHz. Similarly, in the lower right of Fig. 5(b), the radial distributions of magnetic field along the z-axis are depicted. It is evident that at this resonance frequency, the magnetic field enhancement effect is more pronounced on average, about 5 times greater than the previous.

\section{Conclusion}
This work provides an efficient approach to design a high-Q metasurface decoupling isolator for 1.5T surface coil magnetic resonance imaging. With the multilayer construction, it is possible to ensure a homogeneous field distribution and circular polarization conversion, not only eliminating the coupling between surface coil and metamaterial, but also achieving high-Q value by taking advantage of BIC for extremely high field MRI. The promising results from the study of the decoupling isolator indicates the possibility of applications to all clinical sequences and much higher SNR enhancement and resolution. The frequency shift problem in MRI imaging at 1.5T is expected to be automatically tuned by artificial intelligence in the future, which provides a new direction for the research of MRI in the field of artificial intelligence.\textsuperscript{\cite{45,46}}

\medskip
\textbf{Supporting Information} \par 

The initial design consists of a single dielectric medium with a metal topological structure covering its front and back. The initial metal topological structure featured a serpentine winding with 4 turns. However, based on various simulation results, it became evident that the single-layer configuration had limitations in working within the MHz band, failing to achieve decoupling and magnetic resonance enhancement.
Consequently, we include multiple layers of cylindrical medium, ultimately settling on a 5-level linked structure. While increasing the number of layers did lead to some enhancement of magnetic resonance within the metamaterial, it remained non-uniform, and significant coupling persisted between the surface coil and the metamaterial. To address these issues, we maintained the number of layers and increased the number of serpentine coils to 22 turns. Following several iterations of simulation optimization with varying size parameters, we determined that the optimal number of turns for the polarization converter was 10. To achieve the desired cross-polarization decoupling effect within the frequency band corresponding to 1.5T magnetic field, we conducted further optimizations to the structure. These optimizations included the addition of a metal strip at the bottom and metal top sheets at the top, connecting the winding structure on both sides of each cylindrical substrate. Additionally, we incorporated a Si substrate composed of a periodically arranged hollow structure. This Si substrate serves to localize electromagnetic wave energy and reduce reflected energy loss.

\medskip
\textbf{Acknowledgements} \par 
This work was financially sponsored by National Natural Science Foundation of China (52227814,12104339), National Key Research and Development Program of China (2023YFC2400208) and Open Fund of State Key Laboratory of Millimeter Wave, Southeast University (K202216).






\end{document}